\documentclass[aps,prapplied,reprint,superscriptaddress]{revtex4-1}

\usepackage{graphicx}
\usepackage{dcolumn}
\usepackage{bm}
\usepackage{epstopdf}
\usepackage{color}

\emergencystretch=\maxdimen
\begin{document}

\preprint{AIP/123-QED}

\title[]{Promoting superconductivity in FeSe films via fine manipulation of crystal lattice}

\author{Zhongpei Feng}
\affiliation{Beijing National Laboratory for Condensed Matter Physics, Institute of Physics, Chinese Academy of Sciences, Beijing 100190, China}
\affiliation{Key Laboratory of Vacuum Physics, School of Physical Sciences, University of Chinese Academy of Sciences, Beijing 100049, China}

\author{Jie Yuan}
\email{yuanjie@iphy.ac.cn}
\affiliation{Beijing National Laboratory for Condensed Matter Physics, Institute of Physics, Chinese Academy of Sciences, Beijing 100190, China}
\affiliation{Key Laboratory of Vacuum Physics, School of Physical Sciences, University of Chinese Academy of Sciences, Beijing 100049, China}

\author{Ge He}
\affiliation{Beijing National Laboratory for Condensed Matter Physics, Institute of Physics, Chinese Academy of Sciences, Beijing 100190, China}
\affiliation{Key Laboratory of Vacuum Physics, School of Physical Sciences, University of Chinese Academy of Sciences, Beijing 100049, China}

\author{Zefeng Lin}
\affiliation{Beijing National Laboratory for Condensed Matter Physics, Institute of Physics, Chinese Academy of Sciences, Beijing 100190, China}
\affiliation{Key Laboratory of Vacuum Physics, School of Physical Sciences, University of Chinese Academy of Sciences, Beijing 100049, China}

\author{Dong Li}
\affiliation{Beijing National Laboratory for Condensed Matter Physics, Institute of Physics, Chinese Academy of Sciences, Beijing 100190, China}
\affiliation{Key Laboratory of Vacuum Physics, School of Physical Sciences, University of Chinese Academy of Sciences, Beijing 100049, China}

\author{Xingyu Jiang}
\affiliation{Beijing National Laboratory for Condensed Matter Physics, Institute of Physics, Chinese Academy of Sciences, Beijing 100190, China}
\affiliation{Key Laboratory of Vacuum Physics, School of Physical Sciences, University of Chinese Academy of Sciences, Beijing 100049, China}

\author{Yulong Huang}
\affiliation{Beijing National Laboratory for Condensed Matter Physics, Institute of Physics, Chinese Academy of Sciences, Beijing 100190, China}
\affiliation{Key Laboratory of Vacuum Physics, School of Physical Sciences, University of Chinese Academy of Sciences, Beijing 100049, China}

\author{Shunli Ni}
\affiliation{Beijing National Laboratory for Condensed Matter Physics, Institute of Physics, Chinese Academy of Sciences, Beijing 100190, China}
\affiliation{Key Laboratory of Vacuum Physics, School of Physical Sciences, University of Chinese Academy of Sciences, Beijing 100049, China}

\author{Jun Li}
\email{junli@nju.edu.cn}
\affiliation{Research Institute of Superconductor Electronics, Nanjing University, Nanjing 210046, China}

\author{Beiyi Zhu}
\affiliation{Beijing National Laboratory for Condensed Matter Physics, Institute of Physics, Chinese Academy of Sciences, Beijing 100190, China}

\author{Xiaoli Dong}
\affiliation{Beijing National Laboratory for Condensed Matter Physics, Institute of Physics, Chinese Academy of Sciences, Beijing 100190, China}
\affiliation{Key Laboratory of Vacuum Physics, School of Physical Sciences, University of Chinese Academy of Sciences, Beijing 100049, China}

\author{Fang Zhou}
\affiliation{Beijing National Laboratory for Condensed Matter Physics, Institute of Physics, Chinese Academy of Sciences, Beijing 100190, China}
\affiliation{Key Laboratory of Vacuum Physics, School of Physical Sciences, University of Chinese Academy of Sciences, Beijing 100049, China}

\author{Huabing Wang}
\affiliation{Research Institute of Superconductor Electronics, Nanjing University, Nanjing 210046, China}

\author{Zhongxian Zhao}
\affiliation{Beijing National Laboratory for Condensed Matter Physics, Institute of Physics, Chinese Academy of Sciences, Beijing 100190, China}
\affiliation{Key Laboratory of Vacuum Physics, School of Physical Sciences, University of Chinese Academy of Sciences, Beijing 100049, China}
\affiliation{Collaborative Innovation Center of Quantum Matter, Beijing 100190, China}

\author{Kui Jin}
\email{kuijin@iphy.ac.cn}
\affiliation{Beijing National Laboratory for Condensed Matter Physics, Institute of Physics, Chinese Academy of Sciences, Beijing 100190, China}
\affiliation{Key Laboratory of Vacuum Physics, School of Physical Sciences, University of Chinese Academy of Sciences, Beijing 100049, China}
\affiliation{Collaborative Innovation Center of Quantum Matter, Beijing 100190, China}

\date{\today}

\begin{abstract}

Stabilized FeSe thin films in ambient pressure with tunable superconductivity would be a promising candidate for superconducting electronic devices yet its superconducting transition temperature ($T_c$) is below 10 K in bulk materials. By carefully controlling the depositions on twelve kinds of substrates using pulsed laser deposition technique, high quality single crystalline FeSe samples were fabricated with full width of half maximum $0.515^{\circ}$ in the rocking curve and clear four-fold symmetry in $\varphi$-scan from x-ray diffractions. The films have a maximum $T_c$ $\sim$ 15 K on the CaF$_2$ substrate and do not show obvious decay in the air for more than half a year. Slightly tuning the stoichiometry of the FeSe targets, the $T_c$ becomes adjustable from 15 to $<$ 2 K with quite narrow transition widths less than 2 K, and shows a positive relation with the out-of-plane (\emph{c}-axis) lattice parameter of the films. However, there is no clear relation between the $T_c$ and the surface atomic distance of the substrates. By reducing the thickness of the films, the $T_c$ decreases and fades away in samples of less than 10 nm, suggesting that the strain effect is not responsible for the enhancement of $T_c$ in our experiments.

\end{abstract}

\pacs{74.70.Xa, 74.78.-w, 74.25Uv}

\maketitle

\section {Introduction}

The Fe-based superconductors have attracted blooming attention for its superconducting nature and promising applications.\cite{Kamihara,Haindl} Among various Fe-based superconductors, $\beta$-FeSe possesses the simplest structure for an anti-PbO-type structure (space group of $P4/nmm$) with stacks of Fe$_2$Se$_2$ layers, but displays the most multifarious physical properties.\cite{Hsu, Paglione} The FeSe bulk crystals exhibit a superconducting transition temperature ($T_c$) of 9 K.\cite{Chen,Kasahara} Under external pressure, the $T_c$ can be enhanced up to 38 K for FeSe, which is attributed to a decrease of anion height from the Fe-square planes, highlighting the impact of crystal lattice on superconductivity.\cite{Chen,Sun,Medvedev} Unexpectedly, the $T_c$ can be raised further for one unit cell (UC) of FeSe on SrTiO$_3$ substrate.\cite{Wang,Liu,Peng,Ge,Lee} Because of the extreme sensitivity to oxygen, such ultra-thin films of one or a few UC can only be achieved in high vacuum for in-situ fabrications and characterizations, which limits the researches and applications. Therefore, more stabilized FeSe films comparable to or even better than the bulk crystals are highly desired for the next generation of superconducting electronic devices.

\begin{figure*}[tb]
\includegraphics[width=1\textwidth]{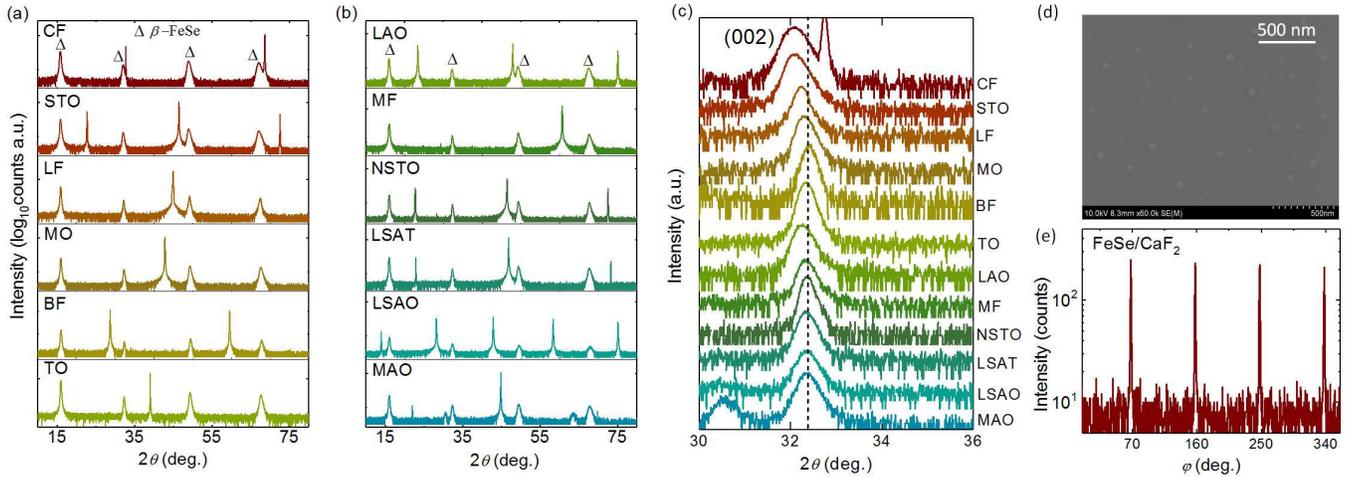}
\caption{(color online). Structure and morphology of FeSe thin films. (a,b) The x-ray 2$\theta$ diffraction patterns for FeSe thin films on various substrates, where the triangular mark demonstrates the Bragg reflection peaks from FeSe crystals. The order of films on various substrates is arranged along $T_c$ dropping, namely, from CaF$_2$ (CF) to SrTiO$_3$ (STO), LiF (LF), MgO (MO), BaF$_2$ (BF), TiO$_2$ (100) (TO), LaAlO$_3$ (LAO), MgF$_2$ (MF), Nb-doped SrTiO$_3$ (NSTO), La$_{0.3}$Sr$_{0.7}$Al$_{0.65}$Ta$_{0.35}$O$_3$ (LSAT), (Sr,La)AlO$_4$ (SLAO), and MgAl$_2$O$_4$ (MAO). (c) The enlarged view of the (002) peaks of the FeSe films. The peaks exhibit obvious shift for different substrates. Here, the dash line is related to the (002) peak of FeSe/MAO. (d) Scanning electron microscope image for a typical FeSe thin film. (e) Azimuthal $\theta$-scan reflection of FeSe/CaF$_2$. A four-fold symmetry is obvious indicating a high-quality epitaxial growth along \emph{c}-axis (00l).}
\end{figure*}

The fabrication of FeSe thin films has been widely studied, and among these researches, pulsed laser deposition (PLD) and molecular beam epitaxy (MBE) techniques are most commonly used.\cite{Haindl,Wang2,Agatsuma,Maeda} From the application point of view, PLD is much more efficient for the growth of films with moderate thickness (above 100 nm).\cite{Agatsuma,Maeda,Han,Chen2,Jung,Nie} Basically, the FeSe can be grown onto various substrates, such as LaAlO$_3$, SrTiO$_3$ and MgO, but their $T_c$ values are generally equal to or lower than that of bulk crystals. A recent report showed that another substrate, CaF$_2$, could enhance the $T_c$ up to 11.4 K with thickness of 150 nm, considerably higher than that of bulk FeSe, which was attributed to the in-plane compressive strain from the substrate.\cite{Nabeshima} The strain by mismatch between the substrate and the film may play a role in promoting the $T_c$  but usually takes effect within limited film thickness, plausibly for the ultra-thin FeSe films where the $T_c$ decreases quickly from 1 UC to 3 UC. \cite{Tan} Therefore, it is still an open question why the $T_c$ is enhanced in thick films like 160 nm. Besides the strain induced by the lattice mismatch,\cite{Nie,Nabeshima,Song} other effects, such as modification of the out-of-plane lattice parameter,\cite{Medvedev} sample inhomogeneity by Fe vacancies,\cite{McQueen,Fang} as well as the growth conditions,\cite{Agatsuma} could also influence the superconductivity. To uncover the substrate effects on the superconductivity, it is important to systematically study the crystal lattice and superconducting properties of FeSe films on various substrates.

\begin{figure}[tb]
\includegraphics[width=0.9\columnwidth,clip]{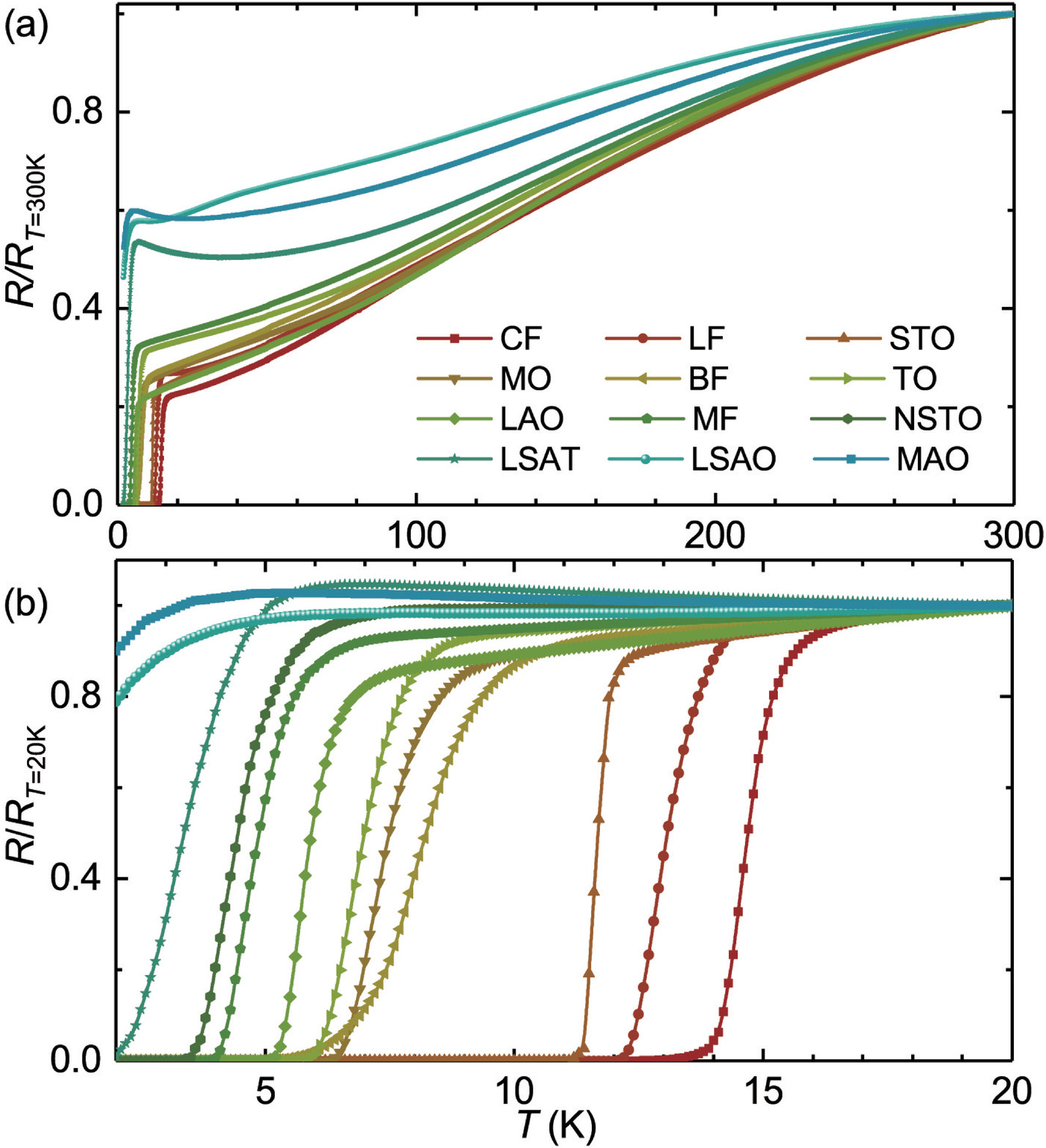}
\caption{(color online). Temperature dependence of normalized resistance $R/R_N$ for FeSe thin films with respect of various substrates, where $R_N$ corresponds to the resistance at 300 K and 20 K, respectively, and the thickness of all films are $\sim$ 160 nm.}
\end{figure}

In this letter, we report the success in synthesizing a series of high-quality single crystalline superconducting FeSe films on various substrates by PLD technique, with maximum $T_{c}$ up to 15 K. Besides, different growth parameters such as the ratio of Fe to Se in the targets and the thickness of the films are also elaborately tuned to arrive at a broad range of zero-resistance transition temperature $T_{c0}$. Based on abundant high quality and stabilized samples, the relation between the crystal lattice and the tunable $T_{c0}$ has been carefully studied.

\section{Experimental}

FeSe polycrystalline targets were fabricated by the solid-state reaction method. The original materials of Fe (4N, Alfa Aesar Inc.) and Se (5N, Alfa Aesar Inc.) powders were mixed with designed ratio of stoichiometry, then heat-treated at 420 $^{\circ}$C for 24 hours in evacuated quartz tubes. The as-prepared material was grinded and sintered at 450 $^{\circ}$C for 48 hours, and such process was repeated more than three times for final targets. FeSe thin films were prepared by PLD technique with a KrF laser. The background vacuum of the deposition chamber is better than 10$^{-7}$ Torr. The FeSe thin films were grown in vacuum with the target-substrate distance of $\sim$ 50 mm, the laser repetition of 2 Hz, and the substrate temperature of 350 $^{\circ}$C.

X-ray diffraction (XRD) measurements of the thin films were performed on a Rigaku SmartLab (9 kW) X-ray diffractometer with Ge(220) $\times$ 2 crystal monochromator. Figure 1(a) and (b) show the XRD data for the FeSe thin films grown on various substrates. In order to avoid the possible epitaxial strain from the substrates, the thicknesses of all films are above 160 nm. All XRD patterns show a fine (00\emph{l}) orientated growth. The corresponding (002) peaks demonstrate slightly leftward shift with $T_c$ increasing as shown in Fig. 1(c). The \emph{c}-axis lattice parameters were calculated from $\theta-2\theta$ XRD patterns by Bragg's law, and the results will be discussed in the latter part. The full width at half maximum (FWHM) of the x-ray rocking curve is $0.515^{\circ}$, showing high crystalline quality. In addition, the fine epitaxy of films is demonstrated by a clear four-fold symmetry in $\varphi$ scan pattern, as shown in Fig. 1(e).

\section{Results and discussions}

First, all the films on various substrates were grown with the same thickness for a better comparison. The transport properties of films were measured in the Physical Property Measurement System (PPMS-9 T). Figure 2 shows the $R-T$ curves for these films. Since the fabrication process is identical, the superconductivity seems to strongly depend on the substrates. In most samples, a zero-resistance transition can be observed except the films on the substrates of MAO, LSAO, and LSAT. Instead, a low-$T$ upturn occurs in the $R-T$ curves of films on MAO and LSAO substrates. While, the films on CF, LF and STO are worthy of more attention, where high onset $T_c$ of 15 K, 13 K and 11.5 K are respectively reached. These values are higher than the previous reports on the same substrates.\cite{Wang2,Agatsuma,Maeda,Chen,Jung,Song}

Since the thickness is considered as a key factor for the superconductivity of thin films, especially for the ultra-thin FeSe system,\cite{Wang,Liu,Song} we study the thickness dependence effect of the FeSe/CaF$_2$ films with the highest $T_c$. In Fig. 3(a), we show the temperature dependence of the resistance for the FeSe films with various thickness adjusted only by the counts of laser pulse. The superconductivity still remains as the thickness is reduced to 10 nm (about 18 UC), where $T_{c0}$ = 2 K. In previous work, strong disorder was usually expected to induce a quantum transition in an ultra-thin system, which could completely suppress the superconductivity.\cite{Nabeshima,Schneider} The existence of superconductivity in the present 10 nm film exhibits the high quality with less disorder. Increasing film thickness, the $T_{c0}$ also ascends gradually and the films display a bulk-like behavior when the thickness exceeds 160 nm.

Considering the composition off-stoichiometry for the superconductivity of FeSe films, namely the Fe or Se vacancy,\cite{Chen2} we prepared the targets with subtle adjustment in the nominal ratio of Fe:Se, including 1:1.10, 1:1.05, 1:1.03, 1:1.00, 1.00:0.99, 1.00:0.97, 1:0.95, 1:0.90, and so on. For comparison, the films were grown with the same thickness of $\sim$160 nm and on the same CaF$_2$ substrate. Figure 3(b) shows the $R-T$ curves for films grown with different targets. $T_{c}$ of the films can be well tuned from below 2 K to 15 K with narrow superconducting transitions ($\Delta T <$ 2 K). It should be noted that the best sample deposited by Fe:Se = 1:0.97 target shows $\Delta T =$ 1.2 K, $R-T$ curve RRR $\sim$ 5, the $T_{c} =$ 15 K, and stable superconductivity for more than half a year, which upgrades the record of the bulk-like FeSe films grown by traditional methods.\cite{Nabeshima}By precise adjustment of target composition, it seems that the ratio of Fe to Se directly determines the superconductivity of the final films. Both energy dispersive x-ray spectra (EDX) and inductively coupled plasma atomic emission spectroscopy (ICP-AES) have been used to check the chemical composition. However, the composition between different superconducting films cannot be clearly defined by these two methods.
\begin{figure}[tb]

\includegraphics[width=0.9\columnwidth,clip]{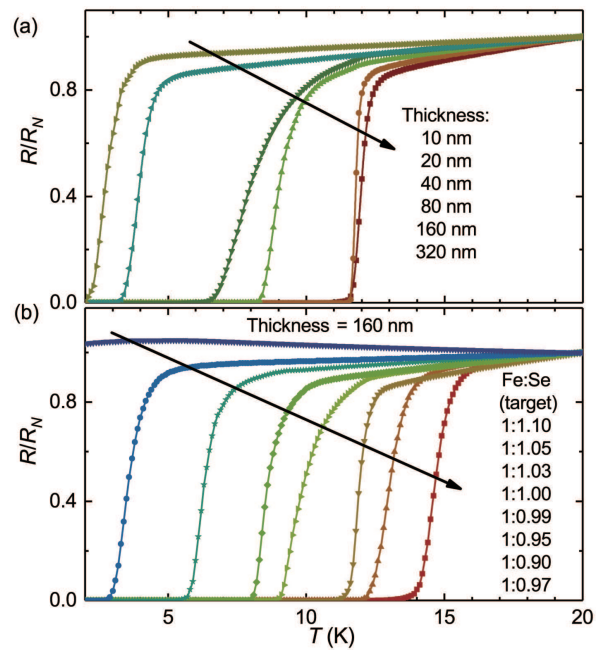}
\caption{(color online). The normalized resistance vs. temperature ($R/R_N-T$) curves for FeSe/CaF$_2$ thin films with respect of thicknesses and targets, here the normal resistance $R_N$ was defined as the resistance at 20 K. (a) A series of FeSe/CaF$_2$ films with various thicknesses are fabricated by adjusting the laser pulsed counts. Here, all films were deposited at 350 $^{\circ}$C by using the same target (Fe:Se = 1:0.97). (b) The FeSe/CaF$_2$ films are grown from different targets with Fe:Se ratio from 1:1.10 to 1:0.90, in which the film deposited from the Fe:Se=1:0.97 target is observed the highest $T_c$.}
\end{figure}

Figure 4(a) gives the zero resistance $T_{c0}$ with respect of the corresponding lattice constant \emph{c} of FeSe films deposited on various substrates. There is an obvious positive correlation between $T_{c0}$ and lattice constant \emph{c} of FeSe films as guided by the dash line, but yet between $T_{c0}$ and the surface atomic distance (\emph{d}) of the substrates (see Fig. 4(b)), indicating the strain from substrate has been relieved in the present bulk-like films. We emphasize that depositing FeSe on the TiO$_2$(100) substrate with rectangular lattice (\emph{b} = 4.593 \AA, \emph{c} = 2.958 \AA) will introduce an anisotropic epitaxial pressure. However, the FeSe/TO film still display superconductivity which is comparable with that of certain films with isotropic strain. Therefore, in comparison with composition for targets, film thickness and substrates, the \emph{c}-axis lattice constant is more closely related to the superconductivity of FeSe. Comparably, the superconductivity of the multi-layered FeSe-based superconductors, i.e., (Li$_{1-x}$Fe$_x$)OHFeSe, are observed as a similar \emph{c}-axis constant dependence behavior,\cite{Dong} which reinforces our understanding on the profile of lattice parameters on the superconductivity.

\begin{figure}[tb]
\includegraphics[width=1\columnwidth,clip]{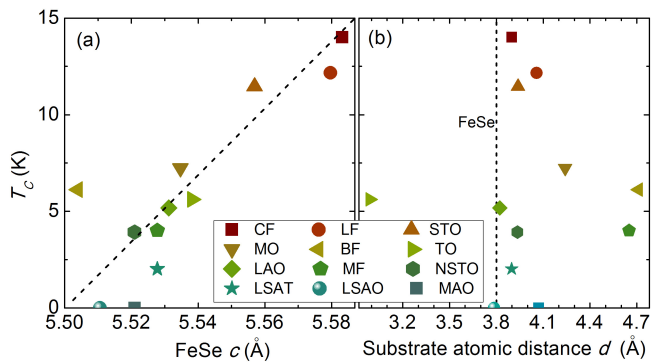}
\caption{(color online). Lattice parameters dependence of superconductivity for FeSe thin films on various substrates. (a) $T_{c0}$ versus $c$ (FeSe lattice constant); (b) $T_{c0}$ versus $d$ (substrate atomic distance at surface).}
\end{figure}

\section{Conclusions}

In conclusion, we have successfully prepared high-quality superconducting FeSe films on such as CaF$_2$, LiF, SrTiO$_3$, LaAlO$_3$, TiO$_2$(100), MgO, BaF$_2$, MgF$_2$, Nb-doped SrTiO$_3$, La$_{0.3}$Sr$_{0.7}$Al$_{0.65}$Ta$_{0.35}$O$_3$, (Sr,La)AlO$_4$ and MgAl$_2$O$_4$. Among them, the films on the CaF$_2$ substrate possess a maximum $T_{c}$ of 15 K, which is the highest value reported so far. By slightly adjusting the ratio of Fe to Se in the targets, a series of FeSe/CaF$_2$ films with tunable $T_{c}$ from $<$ 2 K to 15 K are obtained. The superconductivity of the films on various substrates is found to be mainly dependent on the \emph{c}-axis lattice parameter of FeSe films, in a positive correlation. However, there is no direct correlation between the $T_c$ and the surface atomic distance of substrates, therefore, it is unlikely that the strain effect plays a role in our experiments. The origin of the modification on \emph{c}-axis parameter needs to be further identified, nevertheless, high-quality FeSe thin films with tunable $T_c$  may pave the way for understanding the nature of FeSe from bulk crystal to ultrathin film, and shed light on the applications of superconducting microelectronic devices, such as the hybrid Josephson junctions, single-photon detection superconducting nanowires, and so on.

\begin{acknowledgments}

This work was supported by National Key Basic Research Program of China (Grant Nos. 2015CB921000, and 2016YFA0300301), National Natural Science Foundation of China (Grant Nos. 11674374, 11474338,  11234006, and 61501220), the Key Research Program of Frontier Sciences, CAS (Grant Nos. QYZDY-SSW-SLH001 and QYZDY-SSW-SLH008), Strategic Priority Research Program of CAS (Grant Nos. XDB07020100 and XDB07030200), Beijing Municipal Science and Technology Project (Grant No. Z161100002116011), Jiangsu Provincial Natural Science Fund (No. BK20150561), and Opening Project of Wuhan National High Magnetic Field Center (No. 2015KF19).

\end{acknowledgments}

\end{document}